# Pulsar + escape: the recipe for LHAASO sources ?

Pierrick Martin[1,*], Inès Mertz[1,2], Jean Kempf[1],
Ruo-Yu Liu[5,6] Alexandre Marcowith[4], and Marianne Lemoine-Goumard[3]

[1] IRAP, Université de Toulouse, CNRS, CNES, F-31028 Toulouse, France
[2] Institut supérieur de l'aéronautique et de l'espace, Université de Toulouse, F-31400 Toulouse, France
[3] Université Bordeaux, CNRS, LP2I Bordeaux, UMR 5797, F-33170 Gradignan, France
[4] LUPM, Université de Montpellier, CNRS/IN2P3, CC72, Place Eugène Bataillon, F-34095 Montpellier Cedex 5, France
[5] School of Astronomy and Space Science, Nanjing University, 163 Xianlin Avenue, Nanjing 210023, People's Republic of China
[6] Key Laboratory of Modern Astronomy and Astrophysics, Nanjing University, Ministry of Education, Nanjing 210023, People's Republic of China

**ABSTRACT**

*Context.* Observations of our Galaxy at very-high and ultra-high photon energies have revealed a rich population of sources, many of which have significant angular extension and/or are positionally coincident with energetic pulsars.
*Aims.* We assessed whether the properties of these gamma-ray emitters are consistent with the predictions of a model for the non-thermal leptonic emission of a pulsar wind nebula expanding within a supernova remnant, including the possibility of particle escape across the different components of the system up to the ambient medium.
*Methods.* We used a multi-zone model framework that describes the transport and radiation of a population of ultra-relativistic electron-positron pairs. The dynamics of the system is retrieved from numerical hydrodynamics simulations, which allows to follow the evolution through advanced stages. Model predictions for typical parameter setups are compared to source catalogs issued by the H.E.S.S. and LHAASO collaborations.
*Results.* If particles remain confined in the nebula, $1-100$ TeV emission is only partially consistent with the properties of observed sources in terms of surface brightness, angular extent, and photon index. In particular, the systems spend a long time in a bright and soft state that has no equivalent in observations. Conversely, including the possibility of energy-dependent particle escape across the object results in a much better agreement, both in the $1-10$ TeV range explored with H.E.S.S. or LHAASO-WCDA, and in the $20-100$ TeV range explored by LHAASO-KM2A. Emission components with intermediate to low brightnesses and large to very large sizes result from particles trapped in the remnant or escaped in the surrounding medium.
*Conclusions.* Particle escape, clearly seen in the form of misaligned jets or halos around middle-aged pulsars, is a very important process in much younger systems with $1-100$ kyr ages, and shapes their appearance as TeV/PeV sources.

**Key words.** pulsars: general – cosmic rays – gamma rays: general – astroparticle physics

## 1. Introduction

The development, over the past three decades, of ground-based air-shower arrays as gamma-ray detectors has extended our reach beyond the very-high-energy (VHE) range and definitely paved the way for ultra-high-energy (UHE) astronomy (i.e., involving the detection of > 100 TeV photons; see Cao 2021). Large field-of-view, high duty cycle and increasingly sensitive observatories provided the community with efficient survey capabilities in a fascinating energy window hardly explored up to a few years ago. Building up from pioneering experiments such as CASA-MIA, Tibet AS$\gamma$, MILAGRO, and ARGO-YBG between the late 1980s and early 2000s, the current generation of instruments now routinely enables the detection of UHE photons. The first UHE source reported was the Crab Nebula, using an upgraded version of Tibet AS$\gamma$, with a most energetic photon at $\sim 450$ TeV (Amenomori et al. 2021). This was rapidly followed by the detection of nine sources emitting above 56 TeV using the HAWC observatory, including three extending beyond 100 TeV (Abeysekara et al. 2020). Then, from less than one year of operation with a partly completed array, the LHAASO collaboration achieved the detection of more than 530 UHE photons from twelve sources, including one event at 1.4 PeV (Cao et al. 2021). In this paper, we focus on three particular outcomes of these first few years of exploration of the UHE sky: (i) our Galaxy is populated with numerous UHE sources; (ii) the majority of these sources display significant angular extension; (iii) a connection of these sources with pulsars appears increasingly likely.

The first catalog of LHAASO sources, based on observations from 1 TeV to 2 PeV with the WCDA and KM2A main subsystems, comprises 90 sources within the region of the sky corresponding to declinations between -20° and 80° (Cao et al. 2024). The vast majority of these sources sit in the Galactic plane because photon-photon absorption in the intergalactic medium severely suppresses the flux of extragalactic sources at the highest energies. About a third of the sources may be new, in the sense of lacking previously known TeV source association or exhibiting significantly different morphological properties (the former case being the most common). Seven of the new sources do not have any associations with classical counterparts like supernova remnants (SNRs) or pulsar wind nebulae (PWNe) and are termed dark. Nearly half of the catalog sources are detected above 100 TeV with a significance > 4$\sigma$, and this includes ten sources whose radiation is dominated by emission above a few tens of TeV. The number of LHAASO sources being commensurate with the number of sources found in the H.E.S.S. Galactic Plane Survey (Abdalla et al. 2018b), after just $2-3$ years of operation of LHAASO, this suggests that *most Galactic VHE sources are potential UHE sources* and indicates a *prevalence of PeVatrons in the Galaxy* (quoting from the LHAASO collaboration). The second statement may have to be mitigated somewhat, because the majority of $> 25$ TeV sources have relatively steep spectra with photon indices $\gtrsim 3.0$; so instead of probing PeVatrons, there is a number of cases in which the exquisite sensitivity of KM2A might allow us to probe 100-TeVatrons in a regime where the accelerator runs out of steam (which is already an outstanding achievement providing invaluable information about the physics of acceleration).

---

* pierrick.martin@irap.omp.eu

In the first LHAASO catalog, about 1/3 of the sources are unresolved and point-like for the observatory, while the remaining 2/3 are extended with a 2D Gaussian width $\sigma$ ranging from $0.2°$ to $> 2°$. Assuming sources likely located for the most part between 3 and 6 kpc from us, this translates into physical sizes ranging from $10-20$ pc to $100-200$ pc (for the 39% of the flux contained within $1\sigma$, meaning the full source is at least twice as large). Such extents are only marginally consistent with the size distribution of classical high-energy objects like SNRs or PWNe: under a classical description, an SNR extends up to $20-40$ pc at the time it enters the radiative phase (Cioffi et al. 1988), while a PWN extends up to about $5-15$ pc at most at the time it undergoes reverse-shock crushing (Bandiera et al. 2023b). The physical size implied by LHAASO sources is actually more consistent with the extent over which high-energy objects and/or their progenitors influence a given volume of the interstellar medium (ISM). This can be mediated by fluid processes, when strong outflows like stellar winds or supernova explosions excavate cavities in the ISM (Castor et al. 1975; Weaver et al. 1977), or by kinetic processes, as intense beams of escaping cosmic-rays (CRs) trigger the development of magnetic turbulence (Nava et al. 2016, 2019; Brahimi et al. 2020; Schroer et al. 2021; Plotnikov et al. 2024).

Another surprising development that followed the first looks of the UHE sky is that most sources appear to bear a connection with pulsars, and not so much with SNRs or star clusters that could have been expected to be major source classes as hadronic PeVatrons. Given the modest angular resolution of instruments operating in the VHE/UHE domain, especially ground-based extensive air shower arrays, a deeper analysis of the current observations in conjunction with multi-wavelength information is warranted to formally identify the physical objects behind the sources. That being said, the capability of pulsars to act as leptonic PeVatrons is perfectly established from observations of the Crab nebula (Cao et al. 2021, which does not preclude a small fraction of the energy to go into hadrons), and a subsequent scrutiny of other pulsar-related objects suggests that this capability remains for at least several tens of thousands of years (Sudoh et al. 2021; Breuhaus et al. 2022; de Oña Wilhelmi et al. 2022). That pulsars could power a large number, if not the majority, of VHE/UHE sources was further demonstrated to be viable from energetic, spectral, or population arguments (Fiori et al. 2022; Martin et al. 2022). Cao et al. (2024) mentions that the majority of energetic pulsars in the field-of-view of LHAASO turn out to have an associated LHAASO source. In that respect, the UHE landscape extends and amplifies the trend inferred from VHE observations (Abdalla et al. 2018a).

One possible interpretation of the detection of a plentiful population of UHE sources is that acceleration up to the PeV range is relatively common in our Galaxy and involves powerful objects that are relatively numerous and/or sufficiently long-lived. This is at odds, for instance, with the idea that PeVatrons could proceed from a rare subtype of supernova over a brief phase (Cristofari et al. 2020, although that idea was mostly explored in connection with the acceleration of nuclei). Conversely, this is consistent with a pulsar interpretation, since pulsars are very common objects and remain at the maximum of their spin-down power for durations of $\sim 10^2 - 10^5$ yr. A second implication is that PeV particles have to escape the accelerator swiftly enough to avoid excessive adiabatic and/or synchrotron losses in the intense magnetic fields that were probably needed for efficient confinement and acceleration of the particles. This is consistent with the fact that so many UHE sources are extended and suggests that we might well be probing particles in the early stage of their journey through the Galaxy, rather than accelerators in action. This has a number of important implications for what concerns source identification, since the object that originally energized the particles may have gone weak and less detectable, or non-trivial transport effects may have rendered the connection between the gamma-ray source and the object less obvious (e.g., in mirage sources; see Bao et al. 2024).

In line with these reflections, we explore in this paper the predictions from a recent model framework describing pulsar-powered non-thermal emission in an SNR-PWN system (Martin et al. 2024), and assess to which extent they are consistent with the LHAASO observations reported in the first catalog. The particularity of this framework is to include the possibility of particle escape across the different components of the system and up to the surrounding ambient medium, which seems especially appealing given the widespread observation of extended sources at the highest photon energies. The idea of particle escape from PWNe has been frequently considered in the past in connection to the formation of the local flux of cosmic-ray positrons (Atoyan et al. 1995), a scenario consistent with clear signs that particles are released in the form of jets or halos in relatively old systems (see the review by Olmi 2023). Here, we consider instead particle escape in younger, less dynamically evolved systems, thereby building up on growing observational evidence (e.g., Hinton et al. 2011; Aharonian et al. 2023, 2024). The paper is organized as follows: in Sect. 2, we introduce the model framework and parameter sets used in this work; in Sect. 3, we present the predicted emission in TeV-PeV range and compare it to HGPS and LHAASO catalog data; finally, in Sect. 4, we summarize our findings and discuss certain caveats of the work presented here and a number of possible alternative interpretations of the current observations.

## 2. Modeling TeV-PeV pulsar-powered sources

### 2.1. Theoretical framework

Our work is based on the modeling framework introduced in Martin et al. (2024), the main features of which are briefly recalled here. The model is a one-dimensional spherically symmetric description of an SNR-PWN system, with emphasis on non-thermal leptonic radiation. It assumes that a population of ultra-relativistic electron-positron pairs is produced by the pulsar as it progressively loses energy in a relativistic and magnetized wind and that wind interacts with the surrounding medium. These energetic particles are originally released in the PWN, the bubble of turbulent and magnetized plasma that develops downstream of a pulsar wind termination shock, at the centre of the SNR that was created concurrently to the pulsar in the explosion of the massive star progenitor. The dynamics of the PWN is controlled by the relative evolution of pressure in the nebula (which results from non-thermal energy injected by the pulsar) and pressure in the remnant (which depends on the thermodynamic state of the supersonically expanding stellar ejecta). A distinctive trait of the model is that it implements a two-stage energy-dependent escape process by which particles can diffusively escape the nebula to populate the remnant, and later be decoupled from the remnant and released in the ISM. Along this transport route, particles lose energy as a result of adiabatic and radiative losses, in different proportions depending on evolutionary stage and medium they are in, and the corresponding non-thermal emission from synchrotron radiation and inverse-Compton scattering is computed.

An improved version of this modeling framework is used here and includes the following notable modifications: (i) the prescription for the dynamics of the SNR-PWN system is now extracted from full numerical hydrodynamics simulations performed with the Idefix code (Lesur et al. 2023), instead of using the Sedov-Taylor approximation for the thermodynamic state of the ejecta in advanced stages (Eqs. A9-A10 in Martin et al. 2024); (ii) the magnetic confinement of pairs downstream of the forward shock of the remnant is now prescribed under the assumption of magnetic field amplification from non-resonant instabilities (Bell 2004; Bell et al. 2013), instead of a generic prescription where both the peak value and time decay of the maximum momentum at the shock are free parameters (Eq. 11 in Martin et al. 2024); (iii) accordingly, the average magnetic field in the remnant is now computed under the assumption that the magnetic energy produced in non-resonant instabilities is advected downstream of the forward shock in the SNR volume, instead of using the interstellar magnetic field value (Sect. 2.2.2 in Martin et al. 2024). In practice, improvement (i) allows us to follow the evolution until late times through the so-called reverberation stage (Bandiera et al. 2023a); improvement (ii) is more consistent with the latest theoretical and numerical developments in cosmic-ray acceleration in early-stage SNRs (Caprioli & Spitkovsky 2014; Cristofari et al. 2021); and improvement (iii) yields average magnetic fields in young remnants that are more consistent with the $10 - 100\,\mu G$ values inferred from observations (see for instance Abdo et al. 2010; Ajello et al. 2016; Acero et al. 2022).

We emphasize that the full SNR-PWN model is hybrid in nature and consists in two parts: A) numerical hydrodynamics simulations to compute the dynamics of the whole system; B) multi-zone spectral calculations of the transport and radiation of a population of ultra-relativistic electron-positron pairs. Part A provides the trajectories of many interfaces of the system (forward and reverse shocks, wind termination shock, radius of the nebula) that are subsequently used to define a number of quantities in part B (adiabatic loss rate, diffusive escape time, etc). The consistency of the whole model is ensured by checking that important effects in part B are reflected in part A (this is addressed in Sect. 2.3 when discussing energy losses). Since part A consists of pure hydrodynamical simulations where only thermal effects can be modeled, what it delivers should be considered no more than a hydrodynamical analog of an SNR-PWN system.

### 2.2. Typical parameters of SNR-PWN systems

We aim at demonstrating that typical SNR-PWN systems, described under the framework of Martin et al. (2024) including particle escape, naturally predict VHE/UHE emission whose properties are in agreement with those observed by LHAASO, in terms of surface brightness, source extension, and photon index.

Typical here means involving key parameters that are currently recognized as being the most likely or representative according to a number of studies of pulsars (Faucher-Giguère & Kaspi 2006; Watters & Romani 2011), SNRs (Leahy 2017; Leahy et al. 2020), and PWNe (Torres et al. 2014; Zhu et al. 2023). The corresponding parameter sets for typical SNR-PWN systems are summarized in Table 1 and are organized in four blocks: from top to bottom, pulsar properties, remnant properties, energy injection in the nebula, and particle injection spectrum. Escape-related parameters in the third block are taken from our best-fit to HESS J1809−193 in Martin et al. (2024).

We consider three typical systems, differing from one another by the pulsar spin-down properties. The three sets of initial power and spin-down time scale considered cover the typical span of a population in such a two-parameter plane, with model M2 being representative of the peak of the distribution (see for instance Fig. 1 in Bandiera et al. 2023b). As a starting point, all systems are assumed to be located at a distance of 1 kpc from the Sun at Galactic coordinates $(l, b) = (30.0°, 0.0°)$, but we will later handle intensities or surface brightnesses such that distance is essentially factored out. In addition to the parameters listed in Table 1, the surrounding medium, designated as circumstellar medium (CSM) in the following, is assumed to consist in a magnetic field of typical strength $3\,\mu G$ and a radiation field following the model of Popescu et al. (2017). The latter two assumptions determine the level of radiative losses for leptons that escaped out there.

### 2.3. Dynamics of the SNR-PWN system

Based on the parameters in the top two blocks of Table 1, we simulated the hydrodynamical evolution of a PWN inside an SNR using the Idefix code (Lesur et al. 2023). Idefix is a Godunov-type finite-volume code that we configured such that cell reconstruction is done following a second-order accurate piecewise linear reconstruction using the van Leer slope limiter, while inter-cell fluxes are computed using the HLLC Riemann solver. Time marching is achieved using an explicit multi-stage scheme for which we adopted a second-order Runge-Kutta method, while the time step is determined under a Courant-Friedrichs-Lewy safety factor of 0.5. Under the assumption of spherical symmetry, the one-dimensional spatial grid runs from $r^\star = 10^{-5}$ to 2.5 in characteristic units (approximately spanning 35 pc), and it consists of three consecutive blocks, each with uniform spacing: a first block of 2000 points until position 0.05, followed by a second block of 1500 points until position 0.5, and a third block with 2000 points.

In our simulations, stellar ejecta are initially distributed following a velocity profile that is linear with radius and a density profile that consists of a flat core and a power-law envelope (with an index of 9). Meanwhile, the relativistic pulsar wind is mimicked by the injection of a cold and low-density plasma at the centre of the grid, with a purely radial velocity that is much larger than the outer velocity of the ejecta (by a factor of 10 by default). Each run begins at an age $t_{ini} = 50$ yr from a situation where: (i) cold stellar ejecta freely expanded in a uniform environment and all interactions with the ambient medium up to that point were neglected; (ii) a small PWN started to develop at the center of the grid, and its structure is set from the analytical solution given in Blondin et al. (2001). The full simulation is then run from $t^\star = 0$ to 5 in characteristic units. Actual time is obtained as $t = t_{ini} + t^\star \times t_{ch}$, with a characteristic time unit $t_{ch}$ equal to 6210 yr such that we follow the evolution of our SNR-PWN systems up to 31000 yr.

**Table 1.** Model parameters for three typical SNR-PWN systems

|  | M1 | M2 | M3 |
|---|---|---|---|
| Pulsar spin-down power $\dot{E}_0$ ( erg s$^{-1}$) | $3.0 \times 10^{38}$ | $3.0 \times 10^{37}$ | $3.0 \times 10^{36}$ |
| Pulsar spin-down time scale $\tau_0$ (yr) | $6.0 \times 10^{2}$ | $3.0 \times 10^{3}$ | $3.0 \times 10^{4}$ |
| Pulsar braking index | 3.0 | 3.0 | 3.0 |
| Ejecta energy $E_{\rm ej}$ ( erg) | $10^{51}$ | | |
| Ejecta mass $M_{\rm ej}$ ( M$_\odot$) | 10 | | |
| Circumstellar medium density $n_0$ ( H cm$^{-3}$) | 0.1 | | |
| Turbulence scale $\kappa_T$ | 0.02 | | |
| Turbulent energy injection efficiency $\eta_{\rm T}$ | 0.075 | | |
| Magnetic energy injection efficiency $\eta_{\rm B}$ | 0.025 | | |
| Particle energy injection efficiency $\eta_{\rm e}$ | 0.90 | | |
| Particle injection spectrum low-energy index $\alpha_1$ | 1.5 | | |
| Particle injection spectrum high-energy index $\alpha_2$ | 2.6 | | |
| Particle injection spectrum break energy $E_b$ (GeV) | 500 | | |
| Particle injection spectrum cutoff energy $E_c$ (TeV) | 300 | | |

Thermal radiation losses from the remnant have a marginal impact on the dynamics over the time range explored here and were therefore not included in the results presented thereafter. Non-thermal radiation losses from the nebula are more relevant but currently not implemented, which warrants a specific warning. Synchrotron losses deplete the nebula from its internal energy and this may have an impact on the pace at which it grows before reverse-shock crushing, and on its ability to withstand compression and re-expand after reverse-shock crushing. The effects resulting from including or not non-thermal radiation losses are illustrated in Bandiera et al. (2023b). It appears that neglecting losses does not necessarily have a strong impact on the dynamics of the PWN or on the maximum level of compression reached. This obviously depends on a number of parameters, primarily magnetization and energetics, but a conclusion from Bandiera et al. (2023b) is that the extra compression brought by radiative losses is limited for most of the population. Since our purpose here is to provide global trends valid for systems that are representative of the bulk of the population, neglecting radiative losses may not drastically alter our conclusions.

Conversely, the inclusion of particle escape in our model implies an additional energy loss channel for the nebula that turns out to be significant. We therefore implemented in our numerical hydrodynamics simulations an energy loss term to mimic the internal energy loss from diffusive escape (the term 'mimic' here is important since we aim at incorporating a purely non-thermal process in hydrodynamics simulations where only some thermal processes can be described). At each time step in the preliminary hydrodynamics simulation, some of the parameters of the full model are used to compute a single typical escape time scale (following Eq. 9 in Martin et al. 2024). Then, treating the thermal energy content of the nebula in the hydrodynamics simulation as the equivalent of the non-thermal energy content of an actual nebula, the energy loss rate is computed as the ratio of thermal energy in the nebula to the typical escape time scale derived above. The losses are then applied to the volume of the nebula only, between the wind termination shock and the outer boundary of the shocked wind medium. The effect is notable, as discussed below. This energy loss recipe approximates the full calculation (over the entire particle population and taking all energy-dependent effects into account) by better than a factor of two over most of the time range, and it reproduces the cumulated energy loss after 30 kyr with an accuracy of about 5% (the latter quantity being the relevant one for a proper description of reverse-shock crushing and subsequent behaviour).

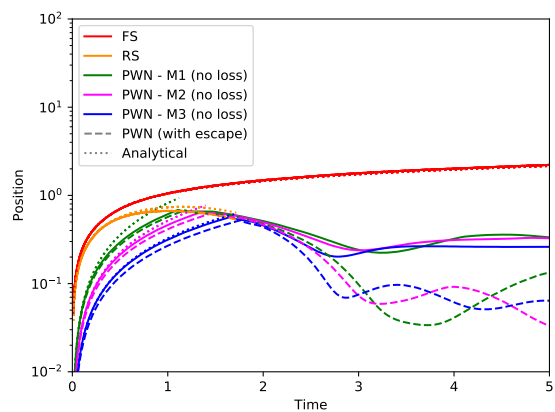

**Fig. 1.** Time evolution of the forward shock (red curves, labelled FS), reverse shock (orange curves, labelled RS), and outer boundary of the PWN (blue curves, labelled PWN). Time and radial position are given in characteristic units. Solid and dashed curves correspond to the output of numerical simulations for the three parameter setups presented in Table 1 (the solutions perfectly overlap for the forward and reverse shocks because SNR parameters are the same in all three cases). The dotted curves correspond to analytical expectations (see text).

The resulting dynamics for the forward and reverse shocks of the SNR and for the outer boundary of the PWN are displayed in Fig. 1. We plot for comparison the analytical formulae from Truelove & McKee (1999) and Blondin et al. (2001) for the SNR and PWN respectively (the curves corresponding to the latter are plotted only for times earlier than reverse-shock crushing of the nebula). A few comments are in order: (i) the mismatch between analytical and numerical results for the SNR are similar to those presented in Truelove & McKee (1999), and due to the necessary simplifications done to produce the formulae; (ii) the mismatch between analytical and numerical results for the PWN are due to the assumption of constant pulsar power in the analytical formula, in contrast to the finite spin-down time scale used here (model M3 for which spin-down power is nearly constant until reverse-shock crushing exhibits a perfect match). Dashed and solid lines for the PWN correspond to the trajectories obtained with and without escape losses, respectively. The effects from such losses are mainly a slower growth in free expansion, and correspondingly a later interaction with the reverse-shock, and a larger level of compression. Without losses, the three models display a maximum compression of the nebula by a factor 2.9/2.6/2.8 in radius, while including losses leads to a stronger compression by a factor 19/18/10 (for models M1/M2/M3, respectively). Losses also seem to attenuate the amplitude of reverberation.

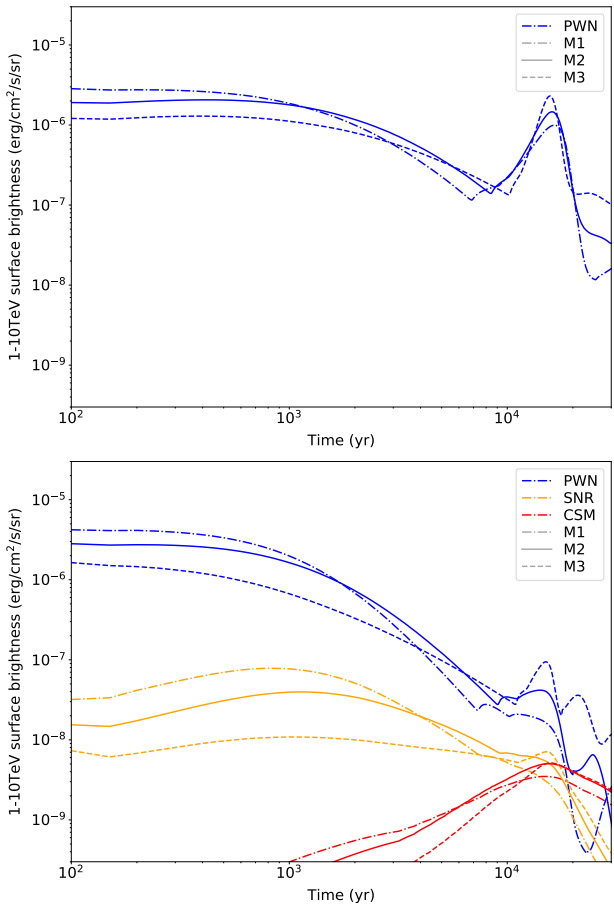

**Fig. 2.** $1-10$ TeV surface brightness as a function of time for our three typical SNR-PWN systems. The top panel results from a calculation without particle escape, and PWN is the only emission component. The bottom panel results from a calculation with particle escape across the system, and gamma-ray emission is contributed by the PWN, SNR, and CSM components altogether. The extent of the CSM component was computed under the assumption of suppressed spatial diffusion, by a factor 1000 with respect to the average Galactic value. Smaller suppression factors would push the set of red curves downwards.

## 3. Predicted TeV-PeV emission

### 3.1. Surface brightness and source size

Based on the dynamical properties introduced above, the evolution of the lepton population in the system and the associated non-thermal emission are computed as described in Martin et al. (2024, with the improvements described in Sect. 2.1). We first illustrate the time evolution of the different emission components and their specific features depending on whether or not particle escape is included. In the following, we will consider surface brightness, computed as the energy flux in a waveband, divided by the solid angle subtended by a given component. The two quantities are computed for our default distance of 1 kpc, but the resulting surface brightness is eventually independent on distance since both the numerator and denominator depend on the inverse of distance squared[1]. For the PWN and SNR components, the size is well defined in our numerical hydrodynamics simulations: forward shock for the former, and outer boundary of the shocked pulsar wind region for the latter. For the CSM component, however, particles that escaped the PWN and subsequently decoupled from the SNR can propagate more or less far away depending on transport conditions in the vicinity of the system. There are indications that a number of high-energy objects are surrounded by a region where spatial diffusion is suppressed with respect to the average Galactic value, by two to three orders of magnitude (see chapter 5 in the review of Tibaldo et al. 2021). Accordingly, we considered by default that diffusion around our SNR-PWN systems is suppressed by three orders or magnitude with respect to the large-scale average value presented in Evoli et al. (2019), and we determined the size of the CSM component as:

$$r_{\text{CSM}}(t, E_e) = \left(r_{\text{SNR}}^2(t) + 4 \times D_{\text{CSM}}(E_e) \times t_{\text{diff}}(E_e)\right)^{0.5} \quad (1)$$

with $D_{\text{CSM}}(E_e) = \chi D_{\text{GAL}}(E_e)$ (2)

and $t_{\text{diff}}(E_e) = \min(t, t_{\text{cool}}(E_e))$ (3)

The typical radius of the CSM component is therefore approximated as the quadratic sum of the radius of the SNR, the minimal size of the component after particle decoupling, and the diffusion length, travelled over the smallest of the system age $t$ and particle cooling time $t_{\text{cool}}$. As mentioned above, diffusion around the system is described by a coefficient $D_{\text{CSM}}$ that is smaller than the large-scale average value $D_{\text{GAL}}$ by a factor $\chi = 10^{-3}$ by default. Because it is partly determined by diffusion, the size of the CSM component depends on a typical particle energy $E_e$, dictated by the gamma-ray band considered. The relation between typical electron energy $E_e$ and mean gamma-ray energy $E_\gamma$ in the inverse Compton scattering process is taken from Aharonian (2004), using for $E_\gamma$ the geometric mean over the band.

Figure 2 displays the $1-10$ TeV surface brightness as a function of time for our three typical SNR-PWN systems, with or without particle escape (bottom and top panel, respectively). In the case without escape, surface brightness is almost constant for a thousand years and subsequently declines because the extent of the source grows faster than its flux. Upon compression by the reverse-shock, particle energization and shrinking of the nebula lead to a surge in surface brightness, followed by a decay during reverberation. When particle escape is included[2], emission from the PWN drops faster, to the point that it is smaller by about an order of magnitude at about 10 kyr, compared to the case without escape. In addition, reverse-shock crushing is accompanied by a much less pronounced rebrightening. The escaped particles feed an emission from the SNR and CSM, at lower surface brightness levels because these components are spatially more extended. At late times $20-30$ kyr, all three components have comparable surface brightnesses, which should then result in relatively extended sources, but the total flux is dominated by far by the SNR and CSM components that are much larger than the PWN.

---

[1] We neglect here the fact that larger distances towards the inner Galaxy would place the object in potentially more intense magnetic and radiation fields.

[2] It might seem surprising that the surface brightness curves for the PWN do not start at the same level at 100 yr, whether or not particle escape is included. This is due to the fact that the magnetic field in the nebula is not exactly the same in both sets of runs. Without escape, 10% of the pulsar power goes into magnetic field. With escape, however, 5% of the power goes into regular magnetic field and another 5% goes into Alfvénic turbulence, and only half of the latter is magnetic energy energy (the other half is kinetic energy).

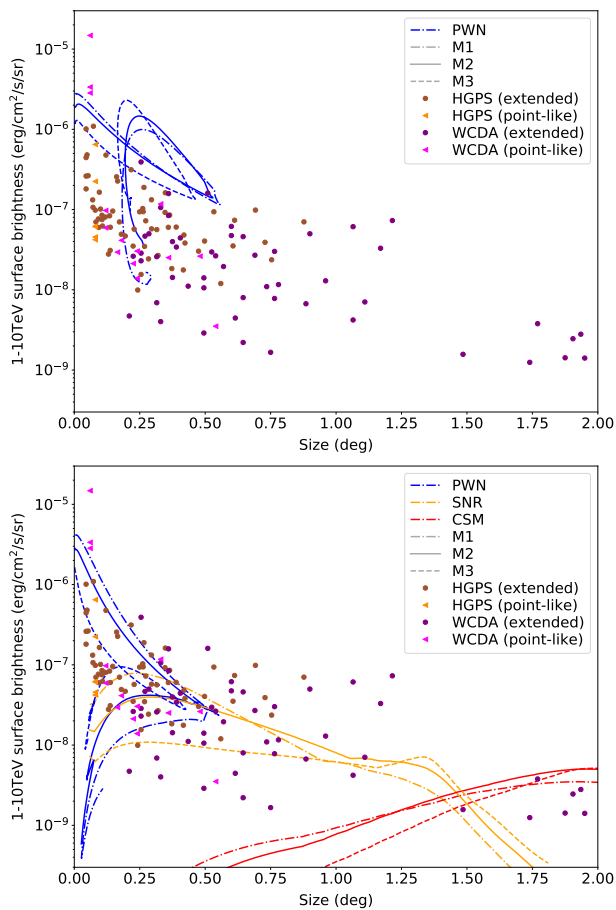

**Fig. 3.** Locus of the time evolution of $1-10$ TeV emission in surface brightness and angular size, for our three typical SNR-PWN systems. The top panel results from a calculation without particle escape, and PWN is the only emission component. The bottom panel results from a calculation with particle escape across the system, and gamma-ray emission is contributed by the PWN, SNR, and CSM components altogether. The extent of the CSM component was computed under the assumption of suppressed spatial diffusion, by a factor 1000 with respect to the average Galactic value. Smaller suppression factors would push the set of red curves down and to the right. The model predictions correspond to a distance of 1 kpc to the system. Larger distances would lead to smaller angular sizes while surface brightnesses would be preserved, thus shrinking all curves towards the ordinate axis. Model-predicted emission is compared to data from the HGPS and LHAASO catalogs.

Figure 3 displays the locus of $1-10$ TeV emission in a surface brightness - angular size plane, for our three typical SNR-PWN systems, with and without particle escape. These model predictions are compared to data from the HGPS and LHAASO catalogs (using WCDA results for the latter). Except HGPS sources with a shell morphology (that are most likely SNRs), we considered all sources from both catalogs irrespective of their identification or classification status. Angular sizes and surface brightnesses correspond to the 68% containment radius. Figure 3 shows that a PWN model without escape for typical remnant and pulsar parameters fails to produce low surface brightnesses and very extended sources, in contrast to what is observed. Conversely, including the possibility of escape yields sources with intermediate brightness and sizes from the SNR component, and sources with low brightness and very large sizes from the CSM component (the latter can be even dimmer and larger if a lower diffusion suppression is considered).

We emphasize a few points that make such a comparison not straightforward. First, the catalog data considered here include objects that may not be described properly within the model framework used here (for instance systems where particle acceleration by the SNR is a dominant component, or pulsar halos from older pulsars like Geminga), or the observations cover only a part of the whole system (for instance Vela X, the full extent of which exceeds the field of view of H.E.S.S.). Second, in the case of HGPS data, continued observations with H.E.S.S. and dedicated analyses revisiting specific sources have led to different emission properties like source extent or photon index (for instance in the case of HESS J1809-193 or HESS J1813-178; see the references in Sect. 1). Third, in the case with escape, the emission at any system age will consist of three components with different brightnesses and sizes and it is not straightforward to anticipate how such an emission layout will be recovered from actual observations and data analysis: the low surface brightness components might fall below detectabilty, or be drown in large-scale interstellar diffuse emission, leaving only the PWN as detectable source; or all three components are detected as one single source with an average size and brightness. In the case of Imaging Atmospheric Cherenkov Telescope systems like H.E.S.S., these aspects are aggravated by the limited field of view of the instrument, which makes it extremely challenging to identify sources with radii beyond $1°$ or so (but there are promising evolutions in analysis techniques; see Wach et al. 2024). The important point is that model runs with escape and typical parameters naturally generates emission components with properties matching those observed in terms of distribution of brightnesses and sizes, while model runs without escape do not.

### 3.2. Photon index

The unsuitability of the model without escape is even more striking when looking at the predicted photon indices. Figure 4 displays the locus of $1-10$ TeV emission in a surface brightness - photon index plane, for our three typical SNR-PWN systems, with and without particle escape. In that plot, which is totally independent of an assumed distance to the source, the SNR and CSM components were grouped together[3]. In the brightness-index plane, the locus of the model-predicted emission is mostly off the distribution of observed sources in cases without escape. The fact that model trajectories (blue curves) partially overlaps the distribution of data (brown dots) is the reason why similar implementations of such a model without escape were successful in accounting for a number of TeV/PeV observations of PWNe, especially with H.E.S.S. (all the more so if one allows for vari-ations from the typical parameters used here). But Fig. 4 shows that the predicted behaviour of the system is characterized by relatively bright and steep TeV emission for most of the lifetime of the system, which is not observed.

---

[3] This is primarily intended to eliminate an artificial spectral feature introduced in the model. By construction, escape from the SNR component is done by clipping the particle distribution above a time-varying maximum energy (see Martin et al. 2024, for details) and transferring the particles to the CSM component. This introduces an abrupt cutoff, and the resulting gamma-ray spectra in the TeV-PeV range can therefore become very steep for the SNR, and very hard for the CSM. Yet, summing the two yields reasonable spectral shapes in agreement with observations, as shown in Fig. 4. It also makes sense because there is substantial overlap in morphology in the SNR and CSM components, such that real observations actually probe the combined emission.

Conversely, in cases including escape, the PWN and SNR+CSM components have predicted spectra whose photon indices are perfectly consistent with observations. For the PWN, the very physical reason behind such a difference is the following. Without escape, particles accumulate in the nebula and, upon compression, they are first energized, which causes a surge in emission, but then rapidly suffer extreme synchrotron losses owing to the enhanced magnetic field; this results in a high flux, because all particles contribute, but steep spectrum, because the bulk of the particle distribution was steepened by energy-dependent losses. With escape, the nebula is efficiently depleted from most of its particle content, so freshly injected particles weigh more in the total population; this results in a low flux, because much fewer particles are left to contribute, but a harder spectrum, because most of the particles were recently injected and had no time to cool.

Figure 5 displays the time evolution of the emission spectra for model setup M2, in two cases with and without article escape and at selected times: during the free expansion phase at 1000 yr, at the time of maximum extent of the nebula when it is hit by the reverse shock at 8000-9000 yr, and when the nebula re-expands after this first compression at 19000-20000 yr. The evolution of the spectra make it easier to make sense of the photon index evolution in Fig. 4. One can see a number of effects like the growing and progressive ageing of the CSM component, or the hardening of the PWN component in the TeV range until reverse-shock crushing and its subsequent softening due to synchrotron burnoff. The two panels also very clearly show that particle escape results in overall smooth spectra at any time, in contrast to the case without escape that leads to very structured spectra after compression (which has never been observed to the best of our knowledge).

Figure 6 shows the locus of $1-10$ TeV and $20-100$ TeV emission in a surface brightness - photon index plane for our baseline model M2 and four variants of it with modified particle injection spectrum: two with a different cutoff energy at 100 and 1000 TeV (instead of 300), and two with a different high-energy index at 2.4 and 2.8 (instead of 2.6). The $20-100$ TeV predictions are compared to data from the LHAASO catalog, more precisely the results from the KM2A subsystem. It can be seen from this plot that some variance in the particle injection spectrum, in agreement with values inferred from studies of individual objects (Torres et al. 2014; Zhu et al. 2023), introduces a spread in photon index and brightness that is consistent with the distribution of observed properties. The impact is particularly noticeable in $20-100$ TeV, showing how sensitive this range is to the high-energy end of the particle injection spectrum. The evolution of the PWN emission in the $20-100$ TeV range is pretty similar to that in $1-10$ TeV (see explanations in the above paragraph). The trend of the SNR+CSM emission is different: it first increases in brightness and hardens, as more and more particles escape the nebula and those have progressively harder spectra, because magnetic field in the nebula and remnant decreases over time and so do synchrotron losses; then, it progressively gets dimmer and steeper as a result of losses in the CSM. Since it is the outcome of a doubly energy-dependent process (diffusion out of the nebula and decoupling from the remnant), the distribution of particles in the CSM is strongly peaked at the highest energies. Over the time frame explored here, its peak shifts from ∼ 100 TeV to ∼ 10 TeV, such that $20-100$ TeV emission fades out to eventually drop abruptly. Overall, in the range explored with LHAASO-KM2A, our model suggests that: (i) high-brightness sources would be accounted for from young systems where the PWN dominate, and a good coverage of current data is obtained from varying primarily the injection cutoff (be it constant or dependent on pulsar power); (ii) low-brightness sources would be accounted for from older systems where the CSM dominate, and a good coverage of the data is obtained from varying the injection index and the level of diffusion suppression (the latter not illustrated here).

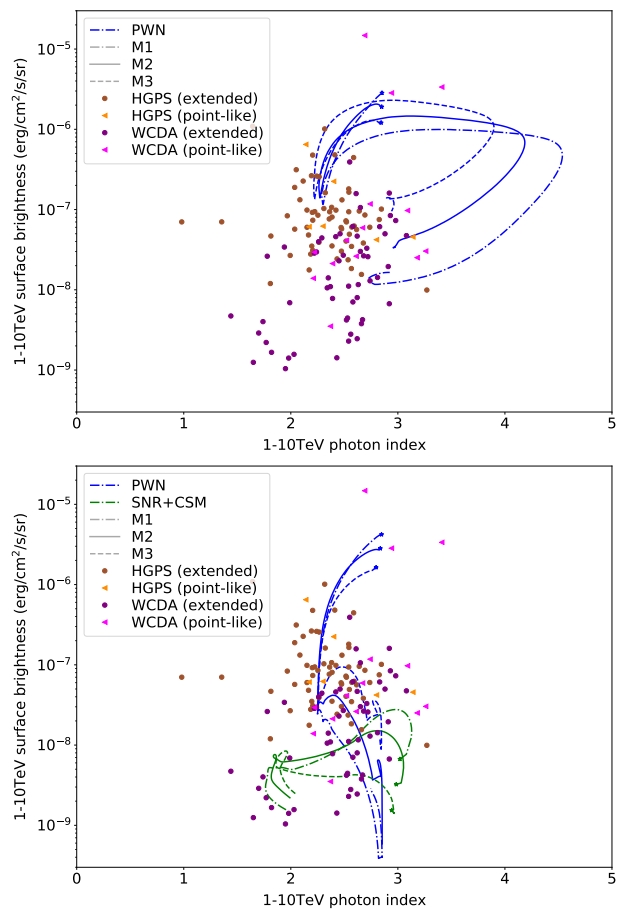

**Fig. 4.** Locus of the time evolution of $1-10$ TeV emission in surface brightness and photon index, for our three typical SNR-PWN systems. The top panel results from a calculation without particle escape, and PWN is the only emission component. The bottom panel results from a calculation with particle escape across the system, and gamma-ray emission is contributed by the PWN, SNR, and CSM components. The extent of the CSM component was computed under the assumption of suppressed spatial diffusion, by a factor 1000 with respect to the average Galactic value, and the SNR and CSM components were grouped together for reasons detailed in the text. Model-predicted emission is compared to data from the HGPS and LHAASO catalogs. Stars indicate the starting point of each track.

## 4. Conclusion

In this work, we explored a possible explanation for the detection with LHAASO of numerous VHE/UHE sources, a large fraction of which have significant angular extension and/or are positionally coincident with energetic non-recycled pulsars. In particular, we worked out predictions from a recent model framework describing non-thermal leptonic emission in an SNR-PWN system.

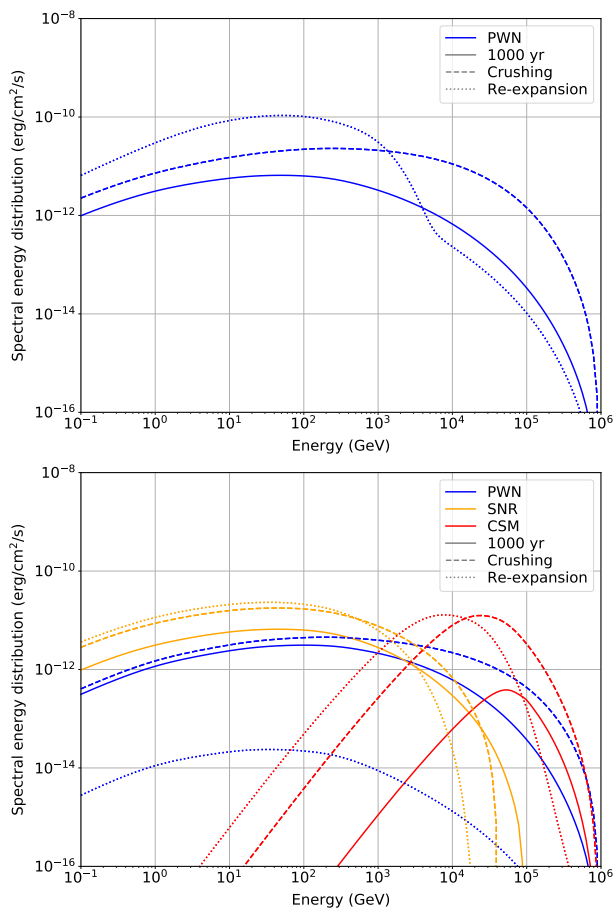

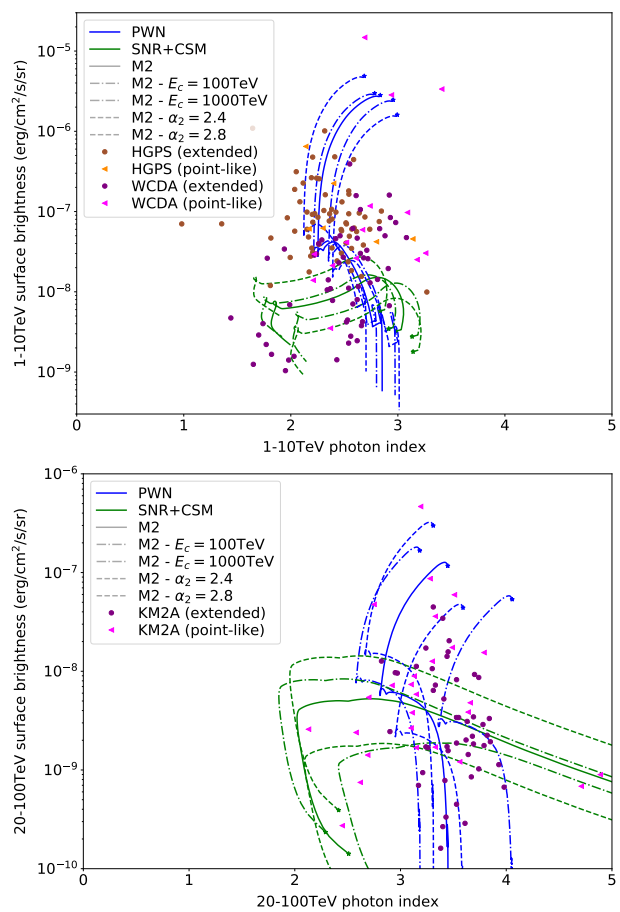

**Fig. 5.** Spectrum of the emission at various times for model M2. The top panel shows the result of a run without particle escape, and emission from the PWN is displayed at 1000, 8000, and 19000 yr (the latter two corresponding to reverse-shock crushing and first re-expansion, respectively). The bottom panel shows the result of a run including particle escape, and emission from the PWN, SNR, and CSM components is displayed at 1000, 9000, and 20000 yr (the latter two still corresponding to reverse-shock crushing and first re-expansion).

**Fig. 6.** Locus of the time evolution of $1-10\,\mathrm{TeV}$ and $20-100\,\mathrm{TeV}$ emission in surface brightness and photon index, for typical SNR-PWN system M2 (see Table 1) and four variants with modified particle injection spectrum. Both panels correspond to a calculation with particle escape across the system. Model-predicted emission is compared to data from the HGPS and LHAASO catalogs (WCDA for the $1-10\,\mathrm{TeV}$ band and KM2A for the $20-100\,\mathrm{TeV}$ band). Stars indicate the starting point of each track.

When used with what is currently understood as the most common parameters for a pulsar-SNR-PWN system, and under the classical assumption that electron-positron pairs remain confined in the nebula, the model predicts a $1-100\,\mathrm{TeV}$ gamma-ray emission over the first 30 kyr of the lifetime of the PWN that is only marginally consistent with the properties of the observed population of VHE/UHE sources listed in the LHAASO and HGPS catalogs. This is true despite the fact that the dynamical evolution of the system is retrieved from numerical hydrodynamics simulations that provide a more reliable description of the system in advanced stages, through the reverberation phase.

Conversely, including the possibility of energy-dependent particle escape from the PWN to the SNR and subsequently to the CSM results in a much better agreement, both in the $1-10\,\mathrm{TeV}$ range explored with H.E.S.S. or LHAASO-WCDA, and in the $20-100\,\mathrm{TeV}$ range explored by LHAASO-KM2A. Particles escaped from the PWN and momentarily trapped in the SNR power emissions with intermediate brightnesses and sizes, while particles decoupled from the SNR and diffusing in the CSM radiate as sources with low brightnesses and large to very large sizes (assuming some level of diffusion suppression in the vicinity of the object).

This work therefore suggests that particle escape is relevant for the dynamical and radiative properties of relatively young SNR-PWN systems with ages up to a few tens of kyr. An observational consequence is that extended gamma-ray halos can be produced around young and powerful pulsars, and not only in the case of middle-aged pulsars like Geminga, although the underlying physics is different.

Yet, the model framework used here is essentially phenomenological and does not directly result from first principles. For that reason, we want to emphasize a number of limitations and possible avenues for improvements:

**Particle escape**: The physical processes involved in particle escape are very likely more intricate than hypothesized here. A better treatment of escape from the PWN into the SNR should include a proper description of magnetic field topology in the full volume of the nebula, including the evolution of turbulence from the inner to the outer boundary and as a function of latitude. Similarly, assessing the escape from the SNR into the CSM would require the full modeling of turbulence inside the volume of the remnant, and a realistic description of the magnetic structure of the forward shock front. Both aspects remain forefront research topics nowadays, as is the investigation of the exact mechanisms responsible for charged particle transport in a magnetized plasma.

**Surrounding environment**: It is desirable to assume a more realistic environment for the evolution of the PWN-SNR system. Instead of the uniform density CSM considered here, a stratified wind-blown bubble is probably more appropriate. The low-density interior of these bubbles may favour the expansion of the PWN and SNR to larger sizes, and constitute an alternative or complementary way to reconcile SNR-PWN models with the observation of (very) extended gamma-ray sources. Such a situation has been considered in other contexts (Meyer et al. 2024), and we plan to update our work along those lines soon.

**Geometry**: A number of effects discussed are strongly tied to the one-dimensional nature of our framework. The compression and reverberation of the nebula are clearly among those. When simulating the evolution of the system in two or three dimensions, hydrodynamical instabilities arise that lead to disruption of the nebula and mixing with the stellar material (Jun 1998; Blondin et al. 2001; Porth et al. 2014). Including the possibility of asymmetries in the problem, like gradients in the ambient density or a non-zero kick velocity for the pulsar, make such a mixing even more complex and vigorous (Kolb et al. 2017). These effects can be seen as alternative or complementary ways to release particles from the PWN into the SNR.

Last, a prime motivation for this study was the detection of very extended sources with LHAASO, an observational fact that may need to be more firmly established. There have been discussions in the literature about the degree of contamination of LHAASO sources by the large-scale interstellar emission of the Galaxy (e.g., Kato et al. 2025). Underestimating the contribution of this component in the mid-plane would lead to overestimating the flux of sources, and maybe also their extent. There is no doubt that continued observations with LHAASO, in conjunction with a better appraisal of diffuse emission in the VHE/UHE range taking into account other information channels like direct cosmic-ray measurements and neutrinos, will help us refine the characterization of these sources over the next few years.

*Acknowledgements.* The authors acknowledge financial support by ANR through the GAMALO project under reference ANR-19-CE31-0014. This work has made use of NASA's Astrophysics Data System Bibliographic Services. The preparation of the figures has made use of the following open-access software tools: Astropy (Astropy Collaboration et al. 2013), Matplotlib (Hunter 2007), NumPy (van der Walt et al. 2011), and SciPy (Virtanen et al. 2020). Pierrick Martin warmly thanks François Rincon for his friendly assistance in setting up and running the Idefix simulations, and Lars Mohrmann for useful comments on an early version of the manuscript. This work was granted access to the HPC resources of CALMIP supercomputing center under the allocation P23022.

## References

Abdalla, H., Abramowski, A., Aharonian, F., et al. 2018a, A&A, 612, A2
Abdalla, H., Abramowski, A., Aharonian, F., et al. 2018b, A&A, 612, A1
Abdo, A. A., Ackermann, M., Ajello, M., et al. 2010, ApJ, 710, L92
Abeysekara, A. U., Albert, A., Alfaro, R., et al. 2020, Phys. Rev. Lett., 124, 021102
Acero, F., Lemoine-Goumard, M., & Ballet, J. 2022, A&A, 660, A129
Aharonian, F., Ait Benkhali, F., Aschersleben, J., et al. 2024, A&A, 686, A149
Aharonian, F., Ait Benkhali, F., Aschersleben, J., et al. 2023, A&A, 672, A103
Aharonian, F. A. 2004, Very high energy cosmic gamma radiation : a crucial window on the extreme Universe
Ajello, M., Baldini, L., Barbiellini, G., et al. 2016, ApJ, 819, 98
Amenomori, M., Bao, Y. W., Bi, X. J., et al. 2021, Phys. Rev. Lett., 126, 141101
Astropy Collaboration, Robitaille, T. P., Tollerud, E. J., et al. 2013, A&A, 558, A33
Atoyan, A. M., Aharonian, F. A., & Völk, H. J. 1995, Phys. Rev. D, 52, 3265
Bandiera, R., Bucciantini, N., Martín, J., Olmi, B., & Torres, D. F. 2023a, MNRAS, 520, 2451
Bandiera, R., Bucciantini, N., Olmi, B., & Torres, D. F. 2023b, MNRAS, 525, 2839
Bao, Y., Giacinti, G., Liu, R.-Y., Zhang, H.-M., & Chen, Y. 2024, arXiv e-prints, arXiv:2407.02478
Bell, A. R. 2004, MNRAS, 353, 550
Bell, A. R., Schure, K. M., Reville, B., & Giacinti, G. 2013, MNRAS, 431, 415
Blondin, J. M., Chevalier, R. A., & Frierson, D. M. 2001, ApJ, 563, 806
Brahimi, L., Marcowith, A., & Ptuskin, V. S. 2020, A&A, 633, A72
Breuhaus, M., Reville, B., & Hinton, J. A. 2022, A&A, 660, A8
Cao, Z. 2021, Universe, 7, 339
Cao, Z., Aharonian, F., An, Q., et al. 2024, ApJS, 271, 25
Cao, Z., Aharonian, F. A., An, Q., et al. 2021, Nature, 594, 33
Caprioli, D. & Spitkovsky, A. 2014, ApJ, 794, 46
Castor, J., McCray, R., & Weaver, R. 1975, ApJ, 200, L107
Cioffi, D. F., McKee, C. F., & Bertschinger, E. 1988, ApJ, 334, 252
Cristofari, P., Blasi, P., & Amato, E. 2020, Astroparticle Physics, 123, 102492
Cristofari, P., Blasi, P., & Caprioli, D. 2021, A&A, 650, A62
de Oña Wilhelmi, E., López-Coto, R., Amato, E., & Aharonian, F. 2022, ApJ, 930, L2
Evoli, C., Aloisio, R., & Blasi, P. 2019, Phys. Rev. D, 99, 103023
Faucher-Giguère, C.-A. & Kaspi, V. M. 2006, ApJ, 643, 332
Fiori, M., Olmi, B., Amato, E., et al. 2022, MNRAS, 511, 1439
Hinton, J. A., Funk, S., Parsons, R. D., & Ohm, S. 2011, ApJ, 743, L7
Hunter, J. D. 2007, Computing in Science Engineering, 9, 90
Jun, B.-I. 1998, ApJ, 499, 282
Kato, S., Alves Batista, R., Anzorena, M., et al. 2025, ApJ, 984, 98
Kolb, C., Blondin, J., Slane, P., & Temim, T. 2017, ApJ, 844, 1
Leahy, D. A. 2017, ApJ, 837, 36
Leahy, D. A., Ranasinghe, S., & Gelowitz, M. 2020, ApJS, 248, 16
Lesur, G. R. J., Baghdadi, S., Wafflard-Fernandez, G., et al. 2023, A&A, 677, A9
Martin, P., de Guillebon, L., Collard, E., et al. 2024, A&A, 690, A116
Martin, P., Tibaldo, L., Marcowith, A., & Abdollahi, S. 2022, A&A, 666, A7
Meyer, D. M. A., Meliani, Z., & Torres, D. F. 2024, A&A, 692, A207
Nava, L., Gabici, S., Marcowith, A., Morlino, G., & Ptuskin, V. S. 2016, MNRAS, 461, 3552
Nava, L., Recchia, S., Gabici, S., et al. 2019, MNRAS, 484, 2684
Olmi, B. 2023, Universe, 9, 402
Plotnikov, I., van Marle, A. J., Guépin, C., Marcowith, A., & Martin, P. 2024, A&A, 688, A134
Popescu, C. C., Yang, R., Tuffs, R. J., et al. 2017, MNRAS, 470, 2539
Porth, O., Komissarov, S. S., & Keppens, R. 2014, MNRAS, 443, 547
Schroer, B., Pezzi, O., Caprioli, D., Haggerty, C., & Blasi, P. 2021, ApJ, 914, L13
Sudoh, T., Linden, T., & Hooper, D. 2021, J. Cosmology Astropart. Phys., 2021, 010
Tibaldo, L., Gaggero, D., & Martin, P. 2021, Universe, 7, 141
Torres, D. F., Cillis, A., Martín, J., & de Oña Wilhelmi, E. 2014, Journal of High Energy Astrophysics, 1, 31
Truelove, J. K. & McKee, C. F. 1999, ApJS, 120, 299
van der Walt, S., Colbert, S. C., & Varoquaux, G. 2011, Computing in Science Engineering, 13, 22
Virtanen, P., Gommers, R., Oliphant, T. E., et al. 2020, Nature Methods, 17, 261
Wach, T., Mitchell, A., & Mohrmann, L. 2024, A&A, 690, A250
Watters, K. P. & Romani, R. W. 2011, ApJ, 727, 123
Weaver, R., McCray, R., Castor, J., Shapiro, P., & Moore, R. 1977, ApJ, 218, 377
Zhu, B.-T., Lu, F.-W., & Zhang, L. 2023, ApJ, 943, 89